\documentclass[3p,twocolumn]{elsarticle}

\usepackage{amssymb}

\usepackage{lineno}
\usepackage{footnote}
\usepackage{hyperref}

\journal{Elsevier}

\begin{document}

\begin{frontmatter}



\title{Novel Silicon n-on-p Edgeless Planar Pixel Sensors for the ATLAS upgrade }


\author[1]{M.~Bomben\corref{cor1}}
\ead{marco.bomben@lpnhe.in2p3.fr}
\author[2]{A.~Bagolini}
\author[2]{M.~Boscardin}
\author[3]{L.~Bosisio}
\author[1,4]{G.~Calderini}
\author[1]{J.~Chauveau}
\author[2]{G.~Giacomini}
\author[5]{A.~La Rosa}
\author[1]{G.~Marchiori}
\author[2]{N.~Zorzi}

\address[1]{Laboratoire de Physique Nucleaire et de Hautes \'Energies (LPNHE)\\  Paris, France\\}
\address[2]{Fondazione Bruno Kessler, Centro per i Materiali e i Microsistemi (FBK-CMM)\\   Povo di Trento (TN), Italy}
\address[3]{Universit\`a di Trieste, Dipartimento di Fisica and INFN, Trieste, Italy}
\address[4]{Dipartimento di Fisica E. Fermi, Universit\`a di Pisa, and INFN Sez. di Pisa, Pisa, Italy}
\address[5]{Section de Physique (DPNC), Universit\'e de Gen\`eve, Gen\`eve, Switzerland}

\cortext[cor1]{corresponding author}

\begin{abstract}
In view of the LHC upgrade phases towards HL-LHC, the ATLAS experiment plans to upgrade the Inner Detector with an all-silicon system. The n-in-p silicon technology is a 
promising candidate for the pixel upgrade thanks to its radiation hardness and cost effectiveness, that allow for enlarging the area instrumented with pixel detectors.
We report on the development of novel n-in-p edgeless planar pixel sensors  fabricated at FBK (Trento, Italy), making use of the Ôactive edgeÕ concept for the reduction of the 
dead area at the periphery of the device. After discussing the sensor technology and fabrication process, we  present device simulations (pre- and post-irradiation) performed 
for different sensor configurations. First preliminary results obtained with the test-structures of the production are shown.
 \end{abstract}


\end{frontmatter}

\linenumbers


\section{Introduction}
\label{sec:intro}

In the next decade the CERN Large Hadron Collider (LHC) will be upgraded  to extend its physics reach;  by 2022 the collider should be capable of a peak 
luminosity of  $10^{35} {\rm cm}^{-2}{\rm s}^{-1}$, the so-called High Luminosity  LHC (HL-LHC)~\cite{HL-LHC} .
By then the ATLAS collaboration will be equipped with a completely new Pixel Detector .  
The innermost layer of the new pixel detector will integrate a fluence of about $10^{16}\, {\rm 1\, MeV\, n_{eq}}/{\rm cm}^2$ for an integrated luminosity of 3000 fb$^{-1}$ 
($\sim$~10 years of operation).
These harsh conditions demand radiation-hard devices and a finely segmented detector to cope with the expected 
high occupancy.

The new pixel sensors will need to have  
  high geometrical acceptance: the future material budget restrictions and  
 tight mechanical constraints require the geometric inefficiency to be less than  2.5\%~\cite{IBL}.
 
In conventional sensor designs there is a relatively large un-instrumented area at the edge of the sensor to prevent the electric field from reaching the rim, where a large
 number of defects are present due to the wafer cutting; 
 for example the current ATLAS pixel sensor has an un-instrumented region of 1.1~mm at the edge~\cite{pixel-electronics-paper}, including Guard Rings (GRs) and providing a 
 suitable  safety margin.  GRs, placed all around the pixel area, can help to improve the voltage-handling capability.

One way to reduce or even eliminate the insensitive region along the device periphery is offered by
 the ``active edge'' technique, in which a deep vertical trench is etched along the device periphery throughout the entire wafer thickness, 
thus performing a damage free cut (this requires using a support wafer, to prevent the individual chips from getting loose). 
The trench is  then heavily doped, extending the ohmic back-contact to the lateral sides of the device: the depletion region can then extend to the edge without causing 
a large current increase.
 This is the technology we have chosen for realizing n-on-p pixel sensors with reduced inactive zone. 

In Section~\ref{sec:active} the active edge technology chosen for a first production of n on p sensors is presented. 
Studies performed with TCAD simulation tools (Section~\ref{sec:simu}) helped in defining the layout  and making a first estimation of the charge collection efficiency 
expected after irradiation. 
In Section~\ref{sec:data} some preliminary results from the electrical characterization of the sensors will be shown.

\section{The active edge sensor production at FBK}
\label{sec:active}

The sensors are fabricated on 100~mm diameter, high resistivity, p-type, Float Zone (FZ), \textless100\textgreater\, oriented, 200~${\rm \mu}$m thick wafers. 
The active edge technology~\cite{bib:Kenney} is used, which is a single sided process, featuring a doped trench, 
extending all the way through the wafer thickness, 
and completely surrounding the sensor. For mechanical reasons, a support wafer is therefore needed, making the back inaccessible after wafer-bonding. 
Several approaches to eventually remove the support wafer are under evaluation; for more details see~\cite{bib:nim2012}. 
After a uniform high-dose boron implant has been performed on the back side, 
the wafers have then been wafer-bonded to a 500~${\rm \mu}$m thick silicon substrate.
Both homogeneous (``p-spray'') and patterned (``p-stop'') implants have been used to insulate the  n-type  pixels; 
the process splittings adopted in the fabrication batch  concern the presence and the doses of these implants.



Two patterned high dose implants are then performed: a  phosphorus implant forming the pixel and GR junctions  and a boron implant for  the ohmic contact
 to the substrate (``bias tab''). 

The etching of the trench is accomplished by a Deep Reactive Ion Etching (DRIE) machine (Alcatel AMS-200), the same used for the fabrication of 
3D detectors~\cite{bib:3DFBK}.

After the trench is etched, its walls are boron-doped in a diffusion furnace. 
Thus, a continuous ohmic contact to the substrate  is created, covering the trench wall and to the backside.
 FBK technology can routinely obtain very uniform, well defined and narrow trenches.

\begin{figure}[!ht]
\begin{center}
\includegraphics[width=0.49\textwidth]{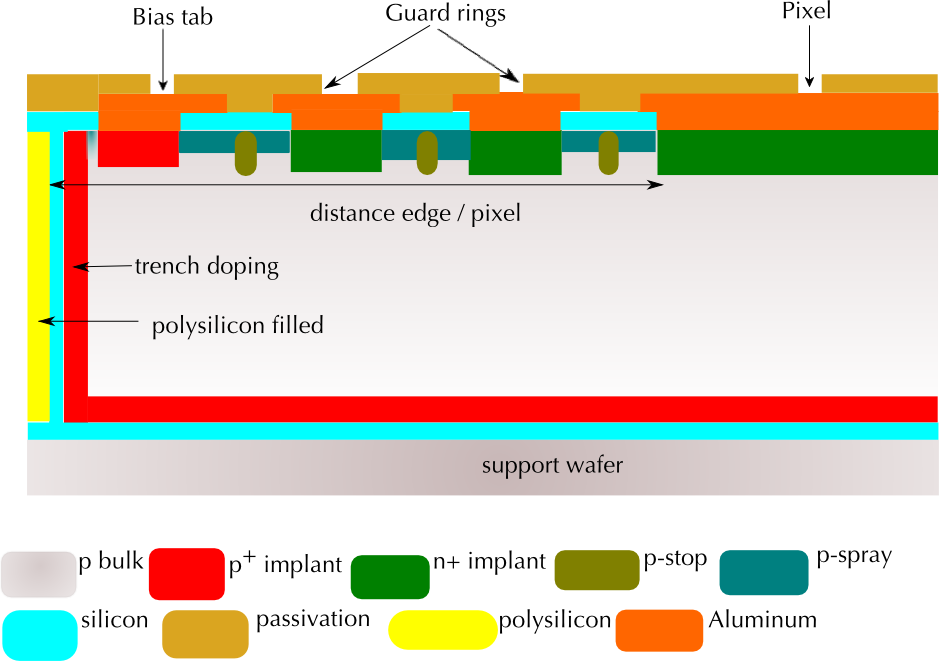}
\caption{\label{fig:pixel}Schematic section of the pixel sensor. The region close to the sensor's edge is portrayed, including the pixel closest to the edge, 
the edge region, including GRs (when present), the bias tab (present only on one edge of the device), the vertical doped trench, and the support wafer.}
\end{center}
\end{figure}

The trenches are then oxidized and filled with polysilicon. 
The remaining processing, arriving at the final device, whose cross-section is sketched in Figure~\ref{fig:pixel}, 
is quite standard, and includes the following steps:  contact opening;  metal deposition and patterning; 
deposition of a passivation layer  (PECVD  oxide)  and patterning of the same in the
 pad and bump-bonding regions.

An additional layer of metal is deposited over the passivation and patterned into stripes, each of them shorting together a row of pixels, contacted through 
 the small passivation openings foreseen for the bump bonding.
This solution has already been adopted for the selection of good 3D FE-I4~\cite{bib:fei4} sensors for the ATLAS IBL~\cite{bib:metal}.
After the automatic current-voltage  measurement 
 on each FE-I4 sensor, the metal will be removed by  wet etching, which does not affect the electrical characteristics  of the devices.

\subsubsection*{Wafer layout}

Nine FE-I4 compatible pixel sensors can be accommodated in a 100~mm wafer.
 The nine FE-I4 sensors differ in the pixel-to-trench distance (100, 200, 300, and 400~${\rm \mu}$m) and in the number of the guard rings (0, 1, 2, 3, 5, and 10)  
surrounding the pixel area (see Figure~\ref{fig:pixel}). The sensor with 3 GRs and a 200~$\mu$m pixel-to-trench distance features two different GR designs, and 
each of them is repeated twice. 
A list of the different FE-I4 sensor versions is reported in Table~\ref{tab:layout_split}.

\begin{savenotes}
\begin{table}[!ht]
\begin{center}
\begin{tabular}{ccc}

Multiplicity & Number of GRs & \begin{tabular}[x]{@{}c@{}}pixel-to-trench\\ distance (${\rm \mu m}$)\end{tabular}   \\
\hline
1 & 0 & 100 \\
1 & 1 & 100 \\ 
1 & 2 &100 \\
4 & 3 & 200 \\ 
1 & 5 & 300 \\ 
1 & 10 & 400  
\end{tabular}
\end{center}
\caption{\label{tab:layout_split}List of FEI4 sensors.  The number of the sensors (first column) is reported for each combination of number of GRs and pixel-to-trench distance. 
Two different designs are envisaged for the sensor with 3 GRs and 200~$\mu$m pixel-to-trench distance. See text for more details.}
\end{table} 
\end{savenotes}

The wafer layout also includes sensors compatible with the FE-I3 read-out chip~\cite{bib:FEI3}, sensors compatible with the OmegaPIX readout chip~\cite{bib:omegapix} and 
many test structures. More details can be found in~\cite{bib:nim2012}.

\section{TCAD simulation}
\label{sec:simu}

In order to explore and compare the properties of the design variations considered, numerical simulations were performed with TCAD tools from SILVACO~\cite{Silvaco}.
 2D structures analogous to the one sketched 
 in Figure~\ref{fig:pixel} have been simulated, varying parameters like the number of GRs and the pixel-to-trench distance. The break down (BD) 
 behaviour of the devices and the charge collection efficiency (CCE) were studied, for simulated un-irradiated and irradiated 
 sensors, with a fluence  $\phi =  1 \times 10^{15} \rm{n_{eq}/cm^2}$; this is the expected fluence for the outer pixel layers  of the new tracker
 at the end of the HL-LHC phase.

 Each of the doped regions (n$^+$ for the pixel and the GRs, p$^+$ for the backside, p-stop, p-spray, bias tab and the trench walls)  
 have been modeled  with simple functions, depending 
 on a set of parameters like the peak concentration and the reference concentration, {\it i.e.} the concentration value at a specified ``rolloff'' distance from the peak position.

Oxide fixed charge density  (with surface density $N_{\rm f}=10^{11}\,{\rm cm^{-2}}$ before irradiation, and $N_{\rm f}=3 \times 10^{12}\,{\rm cm^{-2}}$ after), 
  generation-recombination lifetimes and surface recombination velocity  have been set according to measured IV and CV characteristics of diodes from previous n-on-p 
CiS\footnote{Forschungsinstitut f\"ur Mikrosensorik und Photovoltaik GmbH}  productions. 

The defects at the edge have been modeled with a  1~$\mu$m wide region in which the generation-recombination lifetime was set to a very small value (10$^{-12}$~s; for 
comparison, before irradiation the corresponding value for the bulk is of $10^{-5}$~s). If the  trench doping were not effective, a large current would appear 
as soon as the electric field reaches the edge area.

To describe the radiation damage, an effective model based on three deep levels in the forbidden gap  was used~\cite{bib:Pennicard}. 
Each of these deep levels is defined as 
either donor or acceptor, and is characterized by its energy (with respect to the closest energy band), its capture cross-sections for electrons
($\sigma_e$) and holes ($\sigma_h$) and its 
introduction rate $\eta$, which is the proportionality term between defect concentration and radiation fluence. 

Radiation-induced interface traps at the Si-SiO$_{\rm 2}$ interface are also included in the simulation, as described in~\cite{bib:InterfaceRD50}.


The structure shown in Figure~\ref{fig:pixel} has been slightly modified in the simulations: the support wafer was not present and the backside p$^+$ implant 
was metallized.  
This was done in order to simulate a sensor ready for use. 

The sensors were simulated under reverse bias, applying a negative voltage to the back contact while keeping the pixel at ground potential;
  the 
bias tab was left floating. Different geometries were simulated, varying the number of GRs and the pixel-to-trench distance; see Table~\ref{tab:sim_devices} 
for the list of simulated geometries. If present, the GRs were left floating during the simulations.

\begin{savenotes}
\begin{table}[!ht]
\begin{center}
\begin{tabular}{cc}

\# of GRs & pixel-to-trench distance (${\rm \mu m}$) \\
\hline
 0 & 100 \\
1 & 100 \\ 
2 &100\\
 0 & 200\\ 
 1 & 200\\ 
 2 & 200\\ 
\end{tabular}
\end{center}
\caption{\label{tab:sim_devices}List of simulated sensor layouts.}
\end{table} 
\end{savenotes}

\subsubsection*{Simulation results}

\begin{figure}[!ht]
\begin{center}
\includegraphics[width=0.49\textwidth]{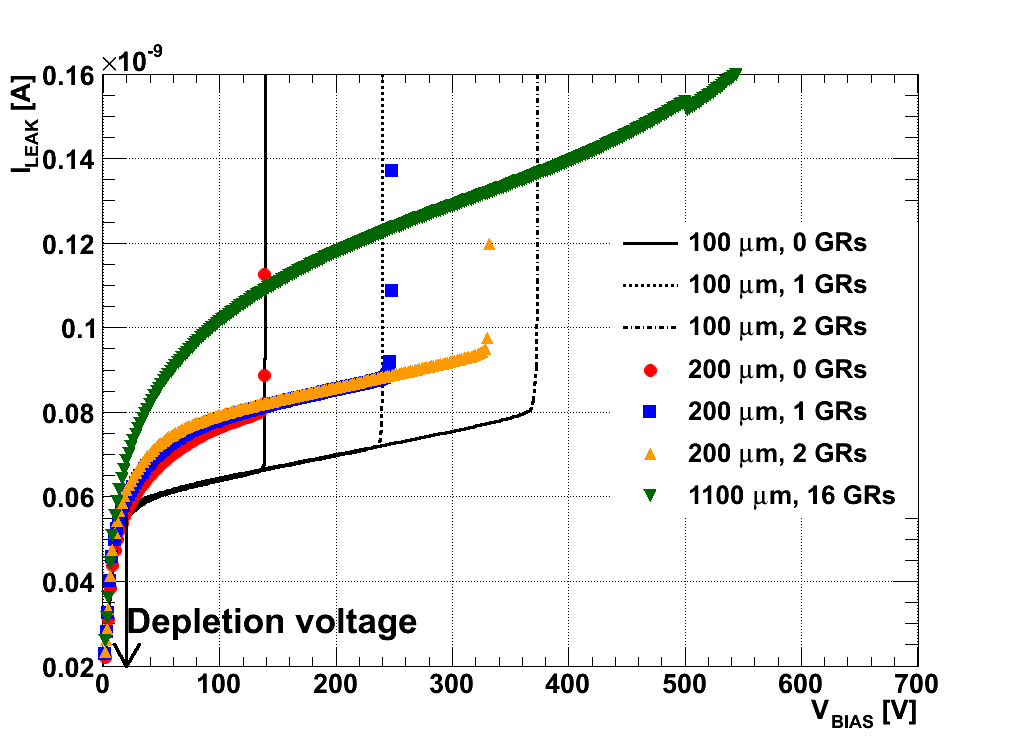}
\caption{\label{fig:BD}Simulated IV curves for the pixel closest to the edge, for several sensor designs before irradiation (see text for details). 
The simulated current has been scaled to reproduce the behaviour of a 50~${\rm \mu}$m wide pixel in the edge direction. 
The depletion voltage is indicated by the arrow.}
\end{center}
\end{figure}

Figure~\ref{fig:BD} shows the current-voltage curves of all the simulated designs, before irradiation. 
The depletion voltage has been estimated using  the AC analysis in the simulations, and determining the depletion voltage value from the fit to the $log(C)- log(V)$ curve; 
 the result was checked against the aforementioned measurements on 
n-on-p diodes from a former production. A sensor with a design compatible with the current ATLAS pixel modules was also simulated; it features a 
pixel-to-trench distance of 1.1~mm and 16 GRs. 
  
From Figure~\ref{fig:BD} it can be seen that before irradiation the BD voltage exceeds by at least 100~V the depletion voltage for all the designs we considered. 
The ATLAS-like sensor shows higher BD voltage with respect to those predicted for our edgeless detectors, 
but all sensors are largely over-depleted before BD. 
Increasing the pixel-to-trench distance yields a higher bulk-generated current, since the depleted volume can further extend laterally. Adding
more GRs 
 greatly helps in increasing the value of BD voltage, extending the operability range of the sensors. The best performance is obtained from a device with 2 GRs 
and a $100 {\rm \mu}m$  pixel-to-trench distance. 

As reported in the literature by different groups ({\it e.g.}~\cite{bib:MPI}), after irradiation the BD voltage increases to much larger values. 
Our simulations of irradiated devices confirm this 
observation.

To study charge collection efficiency (CCE) after irradiation,  charge creation in irradiated sensors was simulated. The most interesting case is when the charge is 
released in the gap between the pixel and the trench, when no GRs are present. If a significant amount of charge can be collected after irradiation in that
 region, the edgeless concept would be verified to work.

Our sensor was illuminated from the front side with a simulated 1060~nm laser beam, setting its power in order to generate the same charge that would be 
released by a minimum ionizing particle (MIP) traversing 200~$\mu$m of silicon ($\sim 2.6$~fC). The laser beam was originating above the front side of the 
detector, with a 2~$\mu$m wide gaussian beamspot. 
 The duty cycle of the laser was  50~ns, with the power ramping up in 1~ns, remaining constant for 10~ns and ramping down in the next nanosecond.

The CCE was studied as a function of the bias voltage for the detector with no GRs and a 100~$\mu$m  trench-to-pixel distance. 
Two incidence points of the laser beam have been considered: one within the pixel and the other in the edge region, at 50~$\mu$m distance from the pixel. 
In the following they will be identified as ``Pixel'' and ``Edge'', respectively.

Based on the properties of the laser beam and of the target material, the simulation program determined the charge of carriers photogenerated inside the device by one pulse. 
The charge collected by the pixel was defined as the integral over the laser duty cycle of the current flowing through the pixel, 
once the stable leakage current had been subtracted. Finally, the CCE was obtained by dividing this collected charge by the total photogenerated charge.

\begin{figure}[h]
\begin{center}
\includegraphics[width=0.49\textwidth]{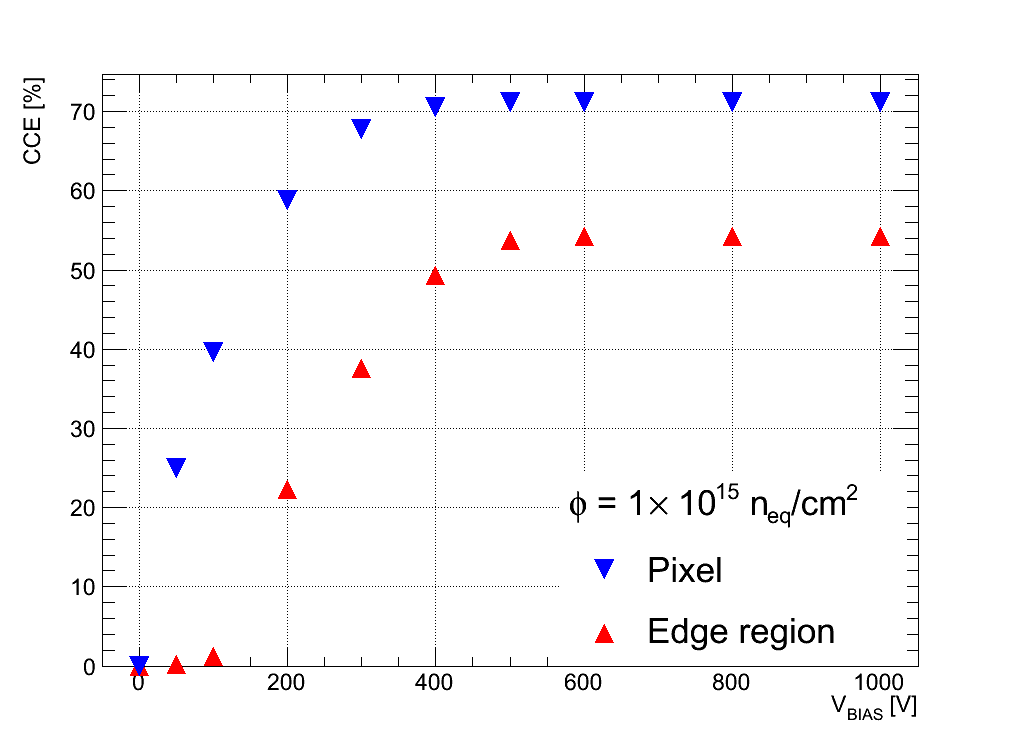}
\caption{\label{fig:CCE_comp_100_0gr}Simulated charge collection efficiency as a function of bias voltage for an irradiated device at a fluence  $\phi= 10^{15} \rm{n_{eq}/cm^2}$ . 
The laser is entering 
the detector either in the pixel region (``Pixel'') or in the un-instrumented region (``Edge region'').
 The sensor has no GRs, and a 100 $\rm{\mu m}$ distance between edge and pixel.}
\end{center}
\end{figure}

In Figure~\ref{fig:CCE_comp_100_0gr} the simulated CCE of an irradiated sensor is presented as a function of the bias voltage  for the two incidence points of the laser beam. 

At a fluence  $\phi = 10^{15} \rm{n_{eq}/cm^2}$  more than 50~\% of the signal  is collected in the ``Edge'' region at a bias voltage of 500~V; 
as a comparison, 70~\% of the  signal is retained in the ``Pixel'' region. 
In both cases the effect of trapping can be observed: the collected charge reaches a {\it plateau} at high voltage, but there the CCE is not of 100~\%. 
No charge is collected from the ``Edge'' region below 100~V: indeed at 100~V bias the electric field is negligible in that region.  
It can be seen that while the maximum CCE for a charge created in the pixel region is reached at a bias voltage above $\sim 400$~V,  in the 
``Edge'' region a bias voltage of 500~V is needed: this is consistent with the depletion zone extending laterally.


Calculations based on trapping time experimental data~\cite{bib:Trapping} for our sensor thickness and  our target fluence produce CCE estimations
  in agreement with our simulations.

\section{First results on real sensors}
\label{sec:data}

\begin{figure}[!h]
\begin{center}
\includegraphics[width=0.49\textwidth]{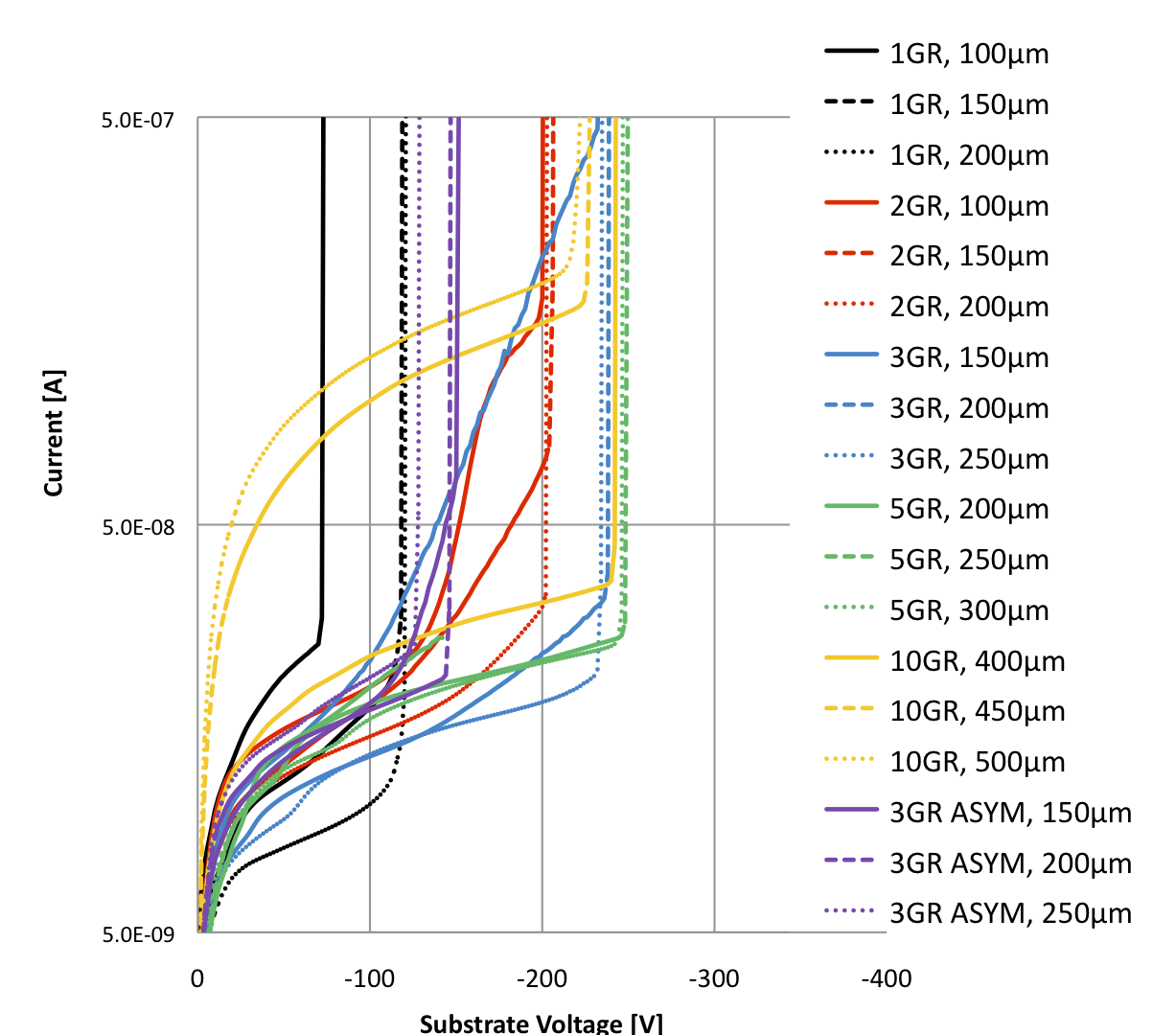}
\includegraphics[width=0.49\textwidth]{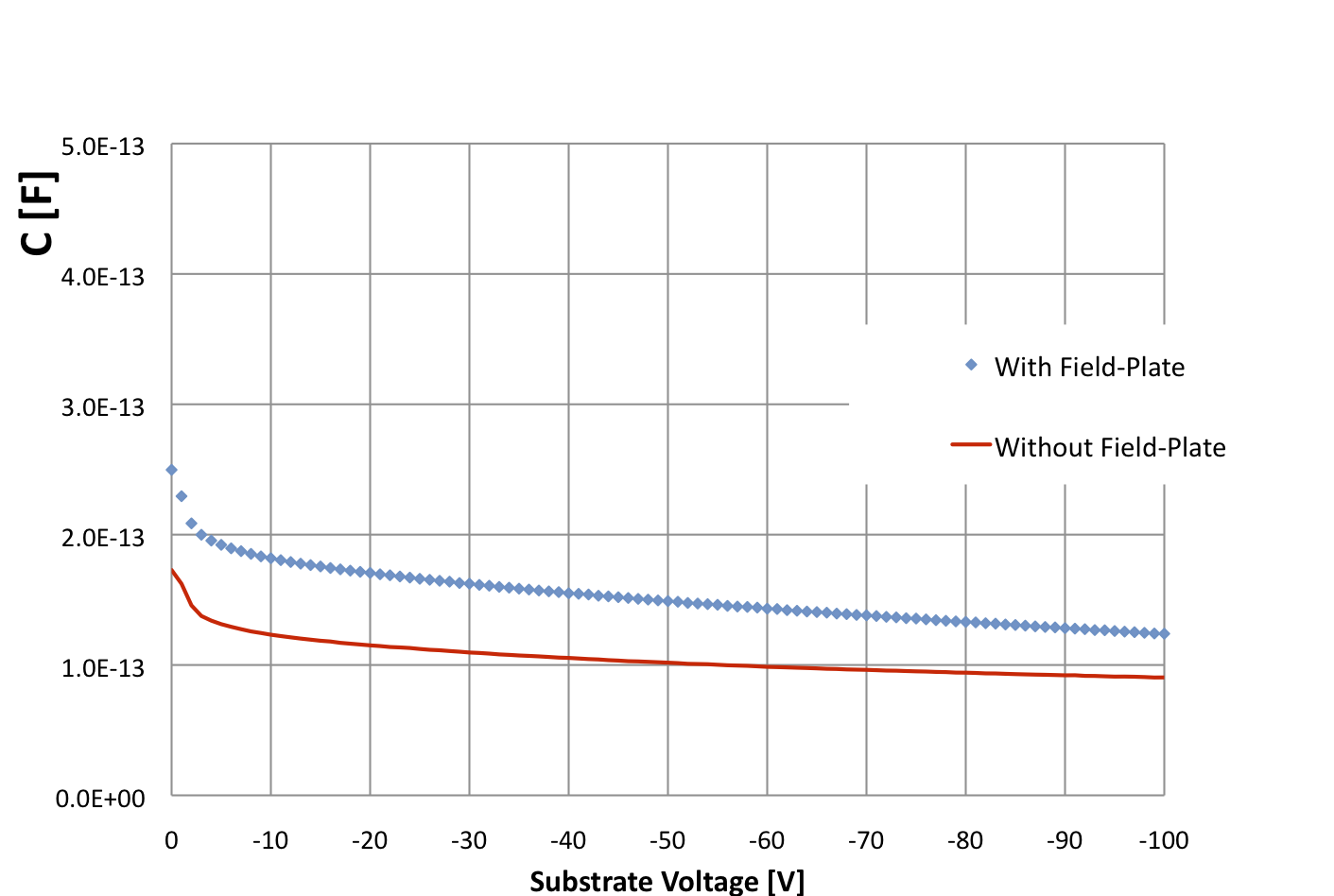}
\caption{\label{fig:iv-testpixels}(Top) IV curves for several test structures, differing by pixel-to-trench distance and by the number of GRs. (Bottom)
 Interpixel capacitance for test structure with FEI4-like cells; the capacitance between the central pixel and
all the other pixels surrounding it in the test structure  is reported as a function of
the bias voltage for pixel cells with a field plate (points), and
without it (solid line).}
\label{default}
\end{center}
\end{figure}

The first wafers have been recently received and the electrical characterization of the production has just started. Test structures consisting of an array of 6~$\times$~30 FE-I4-like
 pixel cells 
have been measured first.  
All the pixels were shorted together and the current voltage characteristics for these sensors is reported in Figure~\ref{fig:iv-testpixels}, top; 
the sensor were inversely polarized via the bias 
tab,  the innermost 
GR was kept at ground (as well as the pixels), and the current flowing through the GR itself is reported. 
As it can be seen, adding more GRs increase the BD voltage and a wider edge-to-pixel distance corresponds to more bulk generated-current; 
all sensors can be operated in over-depletion. 
The simulations reproduce very well these measurements.

For a test structures consisting of an array of 9~$\times$~13 FE-I4-like
 pixel cells, in Figure~\ref{fig:iv-testpixels}, bottom, the capacitance between the central pixel and all the other ones 
 is presented as a function of the bias voltage. It can be seen that the presence of a field-plate increases the interpixel capacitance; the coupling is particularly important due 
 to the presence of the uniform p-spray implant. The level of capacitative coupling, even with  a field-plate, is acceptable in term of electronic noise for the read-out.

\section{Conclusions and outlook}

In view of the upgrade of the ATLAS Inner Detector for HL-LHC runs,
 FBK Trento and LPNHE Paris developed new planar n-on-p pixel sensors, characterized by a reduced inactive region at 
 the edge thanks to a vertical doped lateral surface at the device boundary, the ``active edge'' technology. 
 Simulation studies show the effectiveness of this technique in reducing the dead area, even after 
 simulated fluences comparable to those expected at the end of the HL-LHC phase for the external layers. 

The first, preliminary measurements on real sensors look promising.  Functional tests of the pixel sensors  with radioactive sources 
 and eventually in a beam test, after having bump bonded a number of  pixel sensors to the FE-I4 read out chips, will follow. 

\section*{Acknowledgements} 
We acknowledge the support from the MEMS2 joint project of the Istituto Nazionale di Fisica Nucleare and Fondazione Bruno Kessler






\begin{thebibliography}{99}

\bibitem{HL-LHC}
	M.~Lamont, \href{http://dx.doi.org/10.3204/DESY-PROC-2010-01/8}{\tt http://dx.doi.org/10.3204/DESY-PROC-2010-01/8}



\bibitem{IBL}
	ATLAS TDR 19, CERN/LHCC 2010-013, \\
	\href{http://cdsweb.cern.ch/record/1291633/files/ATLAS-TDR-019.pdf}{\tt http://cdsweb.cern.ch/record/1291633/files/ATLAS-TDR-019.pdf}

\bibitem{pixel-electronics-paper}
	The ATLAS collaboration, JINST 3 P07007, 2008


	


\bibitem{bib:Kenney}
	C.~J.~Kenney {\it et al.}, IEEE Trans. Nucl. Sci. NS-48 (6) (2001) 2405.

\bibitem{bib:nim2012}
	M.~Bomben {\it et al.}, arXiv:1211.5229, submitted to Nucl.\ Instr.\ and Meth. A (Ref.~No.~NIMA-D-12-00985)
	

	
\bibitem{bib:3DFBK}
	C.~Da~Via {\it et al.}, Nucl. Instr. and Meth. A 694 (2012) 321 - 330

\bibitem{bib:fei4}
	M.~Garcia-Sciveres {\it et al.}, Nucl. Instr. and Meth. A  636 (2011) S155-S159.

\bibitem{bib:metal}
	E.~Vianello {\it et al.}, Nuclear Science Symposium and Medical Imaging Conference (NSS/MIC), 2011 IEEE (2011), 523-528


\bibitem{bib:FEI3}
	I.~Peri\'c {\it et al.}, Nucl.\ Instr.\ and Meth. A 565 (2006) 178 - 187.

		


\bibitem{bib:omegapix}
	\textit{\href{http://omega.in2p3.fr/index.php?option=com_content&view=article&id=97&Itemid=240}{OmegaPIX}}


\bibitem{Silvaco}
	\textit{Silvaco, Inc.}\\
4701 Patrick Henry Drive, Bldg 2\\
Santa Clara, CA 95054

\bibitem{bib:Pennicard}
	 D.~Pennicard {\it et al.}, Nucl. Instr. and Meth. A  592 (2008) 16-25.

\bibitem{bib:InterfaceRD50}
	J.~Schwandt {\it et al.}, JINST 7 C01006, 2012

	
\bibitem{bib:MPI}
		P. Weigell {\it et al.}, Nucl. Instr. and Meth. A 658 (2011) 36-40. 
	
\bibitem{bib:Trapping}
	G.~Kramberger {\it et al.}, Nucl. Instr. and Meth. A  476 (2002) 645-651.
	
	
\end{thebibliography}







\end{document}